\documentclass[11pt,a4paper]{article}
\pdfoutput=1
\usepackage{jcappub}

\usepackage[normalem]{ulem}
\usepackage{comment}
\usepackage{ifpdf}
\usepackage{amsfonts}
\usepackage{mathtools}
\usepackage{braket} 
\usepackage{tensor} 
\usepackage{bm} 
\usepackage{slashed}
\usepackage{enumerate}
\usepackage{feynmp}
\usepackage{multirow}
\usepackage{subfigure}
\usepackage{dcolumn}
\usepackage{xcolor}

\begin{document}


\title{Clustering of Primordial Black Holes from QCD Axion Bubbles}
\author[a]{Kentaro Kasai}
\author[a,b]{, Masahiro Kawasaki}
\author[c,d]{, Naoya Kitajima}
\author[a,d]{, \\Kai Murai}
\author[a]{, Shunsuke Neda}
\author[d]{, Fuminobu Takahashi}
\affiliation[a]{ICRR, University of Tokyo, Kashiwa, 277-8582, Japan}
\affiliation[b]{Kavli IPMU (WPI), UTIAS, University of Tokyo, Kashiwa, 277-8583, Japan}
\affiliation[c]{Frontier Research Institute for Interdisciplinary Sciences, Tohoku University, Sendai, 980-8578 Japan}
\affiliation[d]{Department of Physics, Tohoku University, Sendai, 980-8578 Japan}

\abstract{
We study the clustering of primordial black holes (PBHs) and axion miniclusters produced in the model proposed to explain the LIGO/Virgo events or the seeds of the supermassive black holes (SMBHs) in Ref.~\cite{Kitajima:2020kig}.
It is found that this model predicts large isocurvature perturbations due to the clustering of PBHs and axion miniclusters, from which we obtain stringent constraints on the model parameters.
Specifically, for the axion decay constant $f_a=10^{16}~\mathrm{GeV}$, which potentially accounts for the seeds of the SMBHs, the PBH fraction in dark matter should be $f_\mathrm{PBH}\lesssim7\times 10^{-10}$.
Assuming that the mass of PBHs increases by more than a factor of $\mathcal{O}(10)$ due to accretion, this is consistent with the observed abundance of SMBHs.
On the other hand, for $f_a=10^{17}~\mathrm{GeV}$ required to produce PBHs of masses detected in the LIGO/Virgo, the PBH fraction should be $f_\mathrm{PBH}\lesssim6\times 10^{-8}$, 
which may be too small to explain the LIGO/Virgo events, although there is a significant uncertainty in calculating the merger rate in the presence of clustering.
}

\keywords{%
primordial black holes, axions, physics of the early universe
}

\emailAdd{kkasai@icrr.u-tokyo.ac.jp}
\emailAdd{kawasaki@icrr.u-tokyo.ac.jp}
\emailAdd{naoya.kitajima.c2@tohoku.ac.jp}
\emailAdd{kai.murai.e2@tohoku.ac.jp}
\emailAdd{neda@icrr.u-tokyo.ac.jp}
\emailAdd{fumi@tohoku.ac.jp}

\begin{flushright}
    TU-1189
\end{flushright}

\maketitle

\section{Introduction}
\label{sec:intro}

Primordial black holes (PBHs)~\cite{Zeldovich:1967lct, Hawking:1971ei, Carr:1974nx} are black holes formed from overdense regions in the early universe.
Recently, LIGO and Virgo detected the black hole merger events~\cite{LIGOScientific:2016aoc, LIGOScientific:2018mvr, LIGOScientific:2020ibl, LIGOScientific:2021djp}.
The masses of the black holes are estimated to be of $\mathcal{O}(10)\,M_{\odot}$, providing motivation to consider  PBHs in this mass region~\cite{Bird:2016dcv, Clesse:2016vqa, Sasaki:2016jop, Eroshenko:2016hmn, Kashlinsky:2016sdv, Carr:2016drx, Inomata:2016rbd}.
Furthermore, the origin of supermassive black holes (SMBHs) is also a motivation for the existence of PBHs.
SMBHs are black holes with $M \gtrsim 10^{6} M_{\odot}$ and are considered to reside at the centers of galaxies~\cite{Kormendy:1995er}.
SMBHs have been observed at high redshifts $z\gtrsim 6$ (for example, see Refs.~\cite{Banados:2017unc, Wang:2021, Matsuoka:2021jlr, CEERSTeam:2023qgy}).
However, it is difficult for stellar black holes to grow into SMBHs by $z \sim 6$ through the Eddington accretion (for review, see Refs.~\cite{Volonteri:2010wz, Woods:2018lty}).
As an attractive alternative for their origin, PBHs have been studied~\cite{Duechting:2004dk, Kawasaki:2012kn, Nakama:2016kfq, Kawasaki:2019iis, Kasai:2022vhq}.

Recently, PBH formation from inhomogeneous QCD axions was proposed in  Ref.~\cite{Kitajima:2020kig}, motivated by analogous works using the Affleck-Dine mechanism~\cite{Dolgov:2008wu,Blinnikov:2016bxu,Hasegawa:2017jtk,Hasegawa:2018yuy,Kawasaki:2019iis,Kasai:2022vhq}. (See also Refs.~\cite{Li:2023det,Li:2023zyc} for subsequent work on the scenario.)
The QCD axion~\cite{Weinberg:1977ma, Wilczek:1977pj} is a hypothetical particle arising as a (pseudo-)Nambu-Goldstone boson in the Peccei-Quinn (PQ) mechanism~\cite{Peccei:1977hh, Peccei:1977ur}, which solves the strong CP problem.
Assuming that the PQ symmetry is spontaneously broken during inflation, the axion field experiences quantum fluctuations.
As a result, its spatial distribution is characterized by a Gaussian distribution. 
In Ref.~\cite{Kitajima:2020kig}, the spatial distribution of the QCD axion is significantly modified from the Gaussian one by introducing a temporally large PQ breaking term.
Such a temporal PQ breaking term is realized, for example, by the Witten effect~\cite{Witten:1979ey,Fischler:1983sc,Kawasaki:2015lpf,Nomura:2015xil,Kawasaki:2017xwt}, which makes the axion settle down to one of the  minima depending on the inflationary fluctuations. 
Then, depending on the initial position, the universe could be divided into two types of regions with different axion field values.
Subsequently, the potential by the Witten effect disappears by further breaking the residual U(1) symmetry, but the axion field value in each region remains the same.
After that, when non-perturbative QCD effects become relevant, the QCD axion starts to oscillate with different amplitudes in the two different types of regions. 
For a certain choice of parameters, one of them leads to rare and very dense regions where the axion abundance is large, called ``QCD axion bubbles'' or simply ``axion bubbles''. 
After these dense regions enter the horizon, they collapse to PBHs or form miniclusters. 
On the other hand, axions produced in the other regions can account for dark matter.

A fascinating feature of the QCD axion bubble scenario is its capability to produce PBHs and miniclusters within the bubbles, while simultaneously accounting for dark matter. Additionally, there is a one-to-one correspondence between the axion decay constant and the PBH mass. For instance, if the axion decay constant is $\mathcal{O}(10^{16})~\mathrm{GeV}$, the PBH mass becomes $\mathcal{O}(10^4)M_{\odot}$, which overlaps with the mass range required for the seeds of the SMBHs, while if the decay constant is  $\mathcal{O}(10^{17})~\mathrm{GeV}$, the PBH mass becomes $\mathcal{O}(10)M_{\odot}$, which corresponds to the mass accounting for the LIGO/Virgo events.

In this scenario, the properties of the axion bubbles or subsequently formed PBHs depend on the axion field value at the end of inflation.
The perturbations on the bubble scale determine the PBH formation while the perturbations on larger scales control the clustering properties of PBHs. 
The properties of the PBH clustering were recently studied in the PBH formation model using the modified type of Affleck-Dine baryogenesis~\cite{Kawasaki:2021zir} (see also Refs.~\cite{Shinohara:2021psq,Shinohara:2023wjd}), where the clustering was evaluated by considering the two-point correlation function and it was found that the PBH abundance is severely constrained from isocurvature perturbations due to the PBH clustering.
Applying a similar analysis to the QCD axion bubble scenario, one would expect strict limits on the PBH abundance.

\begin{figure}[t!]
    \centering
    \includegraphics[width=.5\textwidth ]{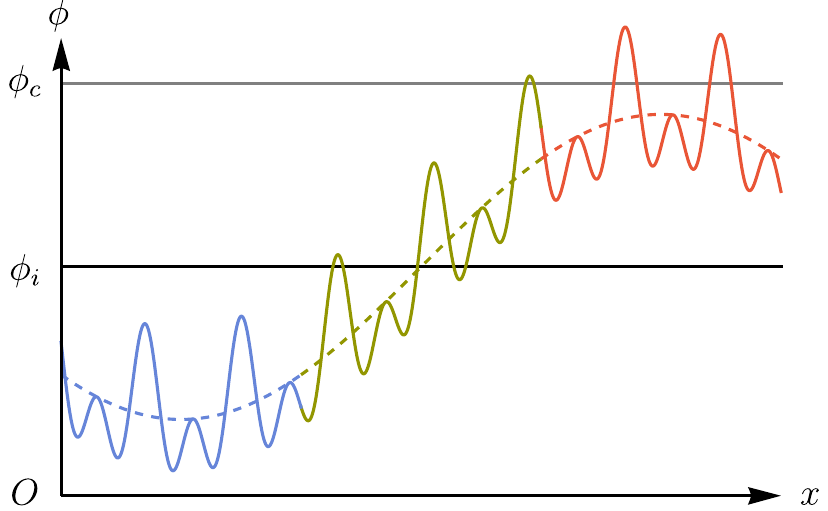}
        \caption{%
        An illustration of the PBH clustering process in this scenario.
        The solid and dashed lines denote larger-scale and smaller-scale perturbations, respectively. 
        If the field value at $x$ is larger than the critical value $\phi_c$, axion bubbles are formed there.
        While the size of bubbles corresponds to smaller-scale perturbations, larger-scale perturbations induce bubble clustering.
        }
    \label{fig: what causes clusterings}
\end{figure}

The clustering of PBHs is induced by large-scale perturbations of the field value.
The schematic image of the PBH clustering is shown in Fig.~\ref{fig: what causes clusterings}.
In this figure, $\phi_i$ is the axion field value averaged over the observable universe, and $\phi_c$ is the threshold field value for bubble formation.
The axion field acquires quantum fluctuations during inflation.
The dashed line denotes larger-scale perturbations, which produce the large-field region (red), the intermediate region (green), and the small-field region (blue).
The solid line denotes the sum of the large-scale perturbations and the smaller-scale perturbations.
The figure shows that more axion bubbles are formed in the large-field region.

The clustering is a characteristic feature of the PBH formation from axion/baryon bubbles, in which perturbations of the field value determines the bubble formation.
In particular, the ordinary scenario in which the PBH formation is due to the inflationary curvature fluctuation does not show the clustering because the density fluctuation is multiplied by $k^2$ to the curvature perturbation and thus the large-scale perturbation (scale larger than the horizon scale at the PBH formation) is suppressed.

In this paper, we evaluate the PBH correlation and isocurvature perturbations in the PBH formation scenario from the axion bubbles.
We derive constraints on the model parameters or the PBH abundance from isocurvature perturbations on the CMB scale.
It is found that the isocurvature perturbations of the axion miniclusters set a severe constraint on this scenario.

The rest of this paper is organized as follows.
In Sec.~\ref{sec: axion_bubble_formation}, we review the scenario of axion bubble formation proposed in Ref.~\cite{Kitajima:2020kig}.
In Sec.~\ref{sec: axion_density}, we also summarize the abundance of PBHs and axion miniclusters estimated in Ref.~\cite{Kitajima:2020kig}.
We calculate the reduced PBH correlation function in Sec.~\ref{sec: correlation_func}.
In Sec.~\ref{sec: isocurvature}, we evaluate the power spectrum of the isocurvature perturbations and obtain the constraints on the axion bubbles.
Finally, Sec.~\ref{sec: summary} is devoted to the conclusion and discussion of our results.

\section{Axion bubble formation}
\label{sec: axion_bubble_formation}

\begin{figure}[t]
    \centering
       \includegraphics[width=0.85\textwidth]{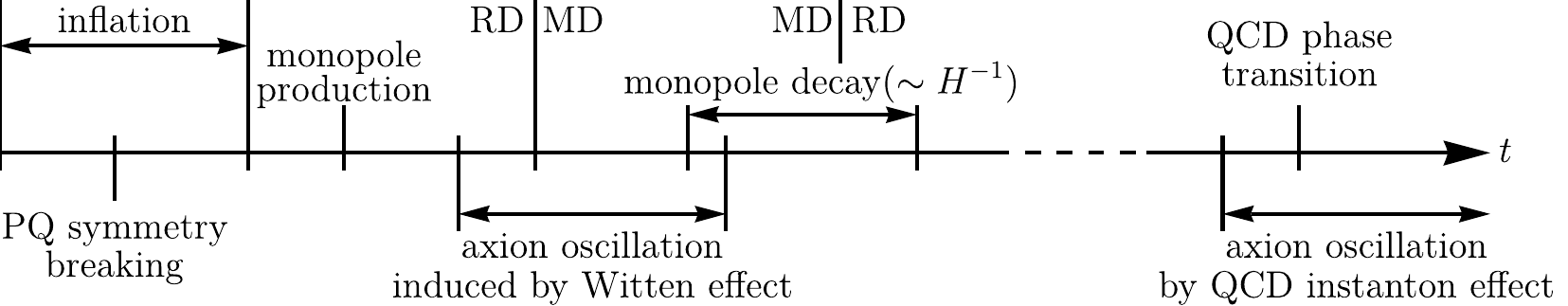}%
    \caption{%
        Time evolution of the universe in this scenario.
    }
    \label{fig: historical scenario}
\end{figure}

In this section, we describe the PBH formation scenario proposed in Ref.~\cite{Kitajima:2020kig}. 
The history of the universe in this scenario is shown in Fig.~\ref{fig: historical scenario}. 
In this scenario, the PQ symmetry is spontaneously broken before or during inflation. Since the QCD axion $\phi$ is massless during inflation, it obtains quantum fluctuations, by which the axion field value has a Gaussian distribution. 

After inflation, $\phi$ acquires two types of potential.
First, the QCD axion couples to gluons through
\begin{align}
    \mathcal{L}
    \supset
    \frac{g^2}{64\pi^2}\frac{\phi}{f_a}\epsilon_{\mu\nu\rho\sigma}F^{a\mu\nu}F^{a\rho\sigma},
\end{align}
where $g$ is the coupling constant of the strong interaction, $f_a$ is the decay constant of the axion, $\epsilon_{\mu \nu \rho \sigma}$ is the totally antisymmetric tensor in four dimensions, and $F^{a\mu\nu}$ is the field strength of gluons. 
Through non-perturbative QCD effects, this term gives the effective potential of the axion field,
\begin{align}
    V_\mathrm{QCD}(\phi)
    =
    m_a^2(T)f_a^2\left(1-\cos{\frac{\phi}{f_a}}\right),
    \label{eq: QCD potential}
\end{align}
where the temperature-dependent mass $m_a(T)$ is given by 
\begin{equation} 
\label{eq: cases f}
    m_a(T)\simeq
    \begin{cases}
        0.57~{\rm{neV}}\left(\frac{10^{16}\,\rm{GeV}}{f_a}\right)\left(\frac{0.15\,{\rm GeV}}{T}\right)^{\frac{c}{2}}  
        &
        T>0.15~{\rm{GeV}}  
    \\[0.5em]
        0.57~{\rm{neV}}\left(\frac{10^{16}\,{\rm GeV}}{f_a}\right)       
        &
        T<0.15~{\rm{GeV}}
    \end{cases}
    \ ,
\end{equation}
with $c = 8.16$~\cite{Borsanyi:2016ksw}.
Here, we choose $\phi = 0$ as the strong CP conserving point.
For later convenience, we define $m_{a0} \equiv m_a(T < 0.15\,\mathrm{GeV})$.

Second, to realize axion bubbles, we assume that $\phi$ also couples to a hidden U(1)$_H$ gauge field through
\begin{align}
    \label{eq:axion_hiddenU1_coupling}
    \mathcal{L} 
    \supset 
    - \frac{\alpha_H}{16 \pi} \left(
        N_H \frac{\phi}{f_a} + \theta_H
    \right)
    \epsilon_{\mu \nu \rho \sigma}
    F_H^{\mu \nu} F_H^{\rho \sigma}
    \ ,
\end{align}
where $\alpha_H$ is the fine-structure constant of the hidden gauge interaction, $N_H$ is an integer called a domain-wall number, $\theta_H$ is the $\theta$-parameter of the U(1)$_H$ gauge symmetry, and $F_H^{\mu \nu}$ is the field strength of the U(1)$_H$ gauge field.
In the following, we take $N_H = 2$.
Note that since we have chosen $\phi = 0$ as the CP conserving point, $\theta_H$ is nonzero in general.

Suppose that some larger symmetry (e.g. SU(2)$_H$) is spontaneously broken into U(1)$_H$ and monopoles are produced after inflation.
In the presence of the coupling~\eqref{eq:axion_hiddenU1_coupling}, U(1)$_H$ monopoles have hidden electric charges through the Witten effect~\cite{Witten:1979ey}. 
The axion field then acquires a mass proportional to $\sqrt{n_M}$ with $n_M$ being the number density of monopoles~\cite{Fischler:1983sc}. 
Since the axion mass decreases more slowly than the Hubble parameter $H$ during the radiation-dominated (RD) era, the axion field starts oscillations when the monopole-induced axion mass becomes comparable to the Hubble parameter.
The monopoles behave as matter, which subsequently leads to the matter-dominated (MD) universe for the parameters of our interest.

\begin{figure}[t]
    \centering
       \includegraphics[width=.625\textwidth ]{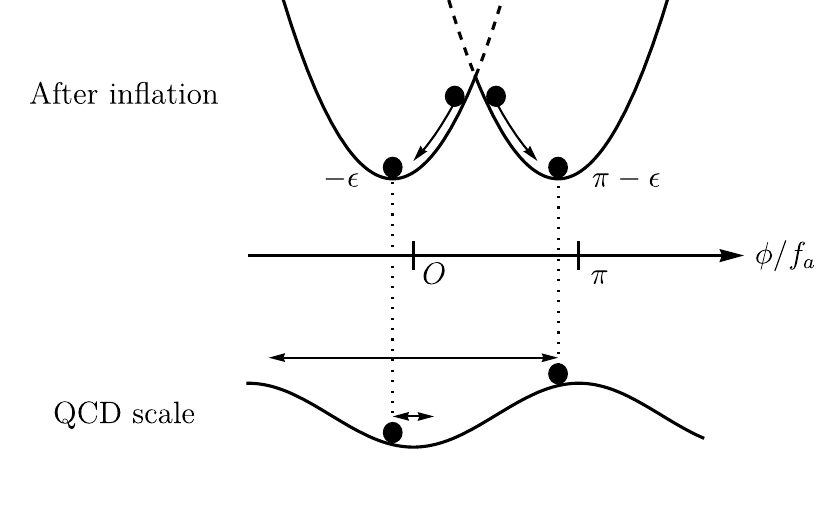}%
    \caption{%
        Potential and dynamics of the axion in the axion bubble formation scenario.
        After inflation, the axion settles in one of the minima of the potential induced by the Witten effect.
        The different minima lead to oscillations with different amplitudes when the QCD potential becomes relevant.
    }
    \label{fig: scenario}
\end{figure}

Since $N_H = 2$, the axion potential caused by the Witten effect has a periodicity of $\pi$ in $\phi/f_a$ and minima at
\begin{align}
    \phi_{\rm min}^{(n)}
    =
    (- \theta_H + 2 \pi n) \frac{f_a}{2}
    \ ,
\end{align}
where $n$ is an integer.
Thus, the axion field rolls down to the nearest minimum depending on its initial value at the end of inflation.
We assume $0 < \theta_H \ll 1$ and define $\epsilon \equiv \theta_H / 2$ so that $\phi_{\rm min}^{(0)}=-\epsilon f_a$.
We depict the potential and dynamics of the axion in Fig.~\ref{fig: scenario}.
Around $\phi_c \equiv (-\epsilon + \pi/2)f_a \simeq \pi f_a/2$, the axion field with $\phi < \phi_c$ ($\phi > \phi_c$) rolls down to $\phi_{\rm min}^{(0)}$ ($\phi_{\rm min}^{(1)}$).
In the following, we assume that $\phi$ rolls down to either of $\phi_{\rm min}^{(0)}$ or $\phi_{\rm min}^{(1)}$ after inflation.\footnote{The potential due to the Witten effect is just an example, and similar results are obtained if there is a potential to the QCD axion due to temporarily large PQ breaking in the early universe~\cite{Takahashi:2015waa}.}

We further assume that the U(1)$_H$ symmetry is spontaneously broken before the QCD phase transition.
As a result, the monopoles decay, and the axion becomes massless again.
Then, around the QCD phase transition,
the axion  acquires a potential $V_{\rm QCD}$ through non-perturbative QCD effects. 
The axion potential from the QCD effect (\ref{eq: QCD potential}) has a periodicity of $2\pi$ in $\phi/f_a$ and takes a minimum value at $\phi = 0$.
Thus, the axion field at $\phi_{\rm min}^{(0)}$ or $\phi_{\rm min}^{(1)}$ starts to oscillate with an initial amplitude $\epsilon$ or $\pi-\epsilon$ around the origin when the axion mass becomes comparable to the Hubble parameter.
Due to the different oscillation amplitudes, the universe is separated into low and high axion density regions.
In this scenario, we set the parameters so that the formation of high-density regions is rare.
Then, regions with high axion density (called axion bubbles) appear in the low axion density background. 
We assume that the background axions account for dark matter and this is achieved by fine-tuning the parameter $\epsilon$ as \cite{Kitajima:2020kig},
\begin{equation}
   \epsilon
   =
   \frac{\phi_{\rm{DM}}}{f_a}
   \simeq
   4.25
   \times 10^{-3}
   \left( \frac{g_{\rm osc}}{60} \right)^{0.209}
   \left( \frac{f_a}{10^{16}\,\rm{GeV}} \right)^{-0.582}
   \ ,
\end{equation}
where $g_{\rm osc}$ is the effective relativistic degrees of freedom at the onset of oscillations due to $V_{\rm QCD}$.
On the other hand, axion bubbles collapse into PBHs if they have sufficiently high energy density at the horizon reentry.
Specifically, if the inside of a bubble is dominated by axions at the horizon reentry, it collapses into a PBH, and otherwise, an axion minicluster is formed.

\section{Axion bubble density}
\label{sec: axion_density}
Here, we discuss the condition for bubble and PBH formation and evaluate the abundance of PBHs and axion miniclusters, following Ref.~\cite{Kitajima:2020kig}.
The axion starts to oscillate at $T \simeq T_{\rm osc}$ satisfying $m_a(T_{\rm osc}) = 3H(T_{\rm osc})$, where $T_{\rm osc}$ is calculated as 
\begin{equation}
   T_{\rm osc}
   =
   0.211~\mathrm{GeV}
   \left( \frac{g_{\rm osc}}{60} \right)^{-0.0822}
   \left(\frac{f_a}{10^{16}~\mathrm{GeV}} \right)^{-0.164}
   \ .
   \label{eq: osc temp}
\end{equation}
At that time, the local number density of axions $n_a$ is given by, 
\begin{equation}
   n_a(T_{\rm osc})
   =
   \frac{1}{2} \kappa m_a(T_{\rm osc}) \phi_{\rm osc}^2 F(\phi_{\rm osc}) \ ,
   \label{eq: axion number density}
\end{equation}
where $\phi_{\rm osc}$ is the axion field value at $T \simeq T_{\rm osc}$, and $\kappa$ is the numerical fudge factor, which is set to $\kappa=1.5$. 
Here, we also include the anharmonic correction factor $F$, which is written as~\cite{Lyth:1991ub,Visinelli:2009zm}
\begin{equation}
   F(\phi_{\rm osc})=\left[1-\log{\left(1-\frac{(\phi_{\rm osc}/f_a)^2}{\pi^2}\right)}\right]^{1.16}
   \ .
   \label{eq: detail definition}
\end{equation}
To evaluate the energy density of the axions, it is convenient to consider the ratio of the axion number density to the entropy density, $n_a/s$.
The entropy density at $T=T_{\rm osc}$ is given by
 \begin{equation}
   s(T_{\rm osc})=\frac{4 \rho_{\rm{rad}}(T_{\rm osc})}{3T_{\rm osc}}
      =  \frac{4H^2(T_{\rm osc})M_\mathrm{Pl}^2 }{T_{\rm osc}}
      =  \frac{4M_{\rm{Pl}}^2m_a^2(T_{\rm osc})}{9T_{\rm osc}} \ ,
   \label{eq: entropy density}
\end{equation}
where $M_{\rm Pl} \simeq 2.4\times 10^{18}$~GeV is the reduced Planck mass.
For $T < T_{\rm osc}$, $n_a/s$ is conserved and thus given by
\begin{equation}
    \frac{n_a(T)}{s(T)}
    =
    \frac{n_a(T_{\rm osc})}{s(T_{\rm osc})}
    =
    \frac{9 \kappa T_{\rm osc}}{8m_a(T_{\rm osc})}
    \left( \frac{\phi_{\rm osc}}{M_{\rm Pl}} \right)^2 F(\phi_{\rm osc})
    \ .
   \label{eq: useful axion abundance}
\end{equation}

\begin{figure}[t]
    \centering
    \includegraphics[width=.75\textwidth ]{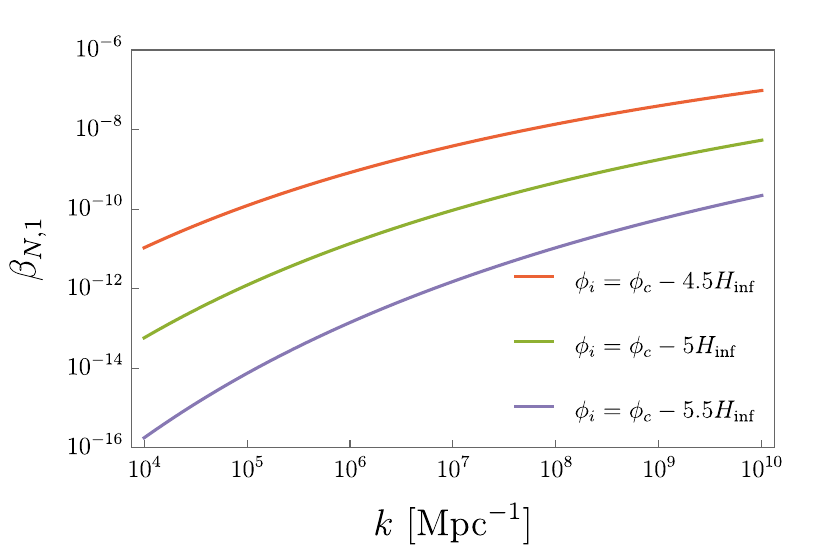}
    \caption{%
        Volume fraction of axion bubbles for different initial conditions $\phi_i$.
        We used Eq.~\eqref{eq: e-fold def} to relate $k$ and $N$.
    }
    \label{fig: volume fraction}
\end{figure}

In the scenario of Ref.~\cite{Kitajima:2020kig}, the bubble formation depends on the axion field value at the end of inflation.
Since the axion is massless during inflation, the axion field diffuses by quantum fluctuations.
The time evolution of the axion field $\phi$ coarse-grained over the horizon scale is described by a probability distribution $P(N,\phi)$, which obeys the Fokker-Planck equation: 
\begin{equation}
    \frac{\partial P(N,\phi)}{\partial N} 
    =
    \frac{H_{\rm{inf}}^2}{8\pi^2}
    \frac{\partial^2 P(N,\phi)}{\partial \phi^2}
    \ ,
    \label{eq: Fokker-Planck}
\end{equation}
where $H_{\rm{inf}}$ is the Hubble parameter during inflation, and $N$ is the e-folding number defined by
\begin{equation}
    N = \ln \frac{a}{a_i}
    \ .
\end{equation}
Here, $a\, (a_i)$ is the (initial) scale factor.
In the following, we assume that $H_\mathrm{inf}$ is constant during inflation.
Then, the solution of this equation is given by
\begin{align}
    \begin{aligned}
        P(N,\phi; \phi_i)
        &=
        \frac{1}{\sqrt{2\pi}\sigma(N)}
        \exp{ \left( -\frac{(\phi-\phi_i)^2}{2\sigma^2(N)} \right) }
        \ ,
        \\
        \sigma(N)
        &=
        \frac{H_{\rm{inf}}}{2\pi}\sqrt{N}
        \ ,
    \end{aligned}
    \label{eq: probability density}
\end{align}
where we set the initial condition $P(0,\phi; \phi_i)=\delta(\phi-\phi_i)$ at the horizon exit of the current horizon scale, $N = 0$. 
If we assume $0 < \phi_i < \phi_c$, the probability that $\phi$ rolls down to the minima other than $\phi_{\rm min}^{(0)}$ or $\phi_{\rm min}^{(1)}$ after inflation is negligibly small.
Thus, we can approximate the total volume fraction of the axion bubbles by 
\begin{align}
    \beta(\phi_i)
    =
    \int_{\phi_c}^\infty P(N_\mathrm{end},\phi; \phi_i)
    {\rm{d}}\phi
    \ ,
\end{align}
where $N_\mathrm{end}$ is the e-folding number at the end of inflation.
The volume fraction of the axion bubbles that exit the horizon between $N$ and $N + {\rm d} N$ is given by $\beta_{N,1}(N;\phi_i) dN$ with
\begin{align}
   \begin{split}
      \beta_{N,1}(N;\phi_i)
      \equiv
      \frac{\partial}{\partial N}
      \int_{\phi_c}^\infty P(N,\phi;\phi_i){\rm{d}}\phi
      =
      \frac{\phi_c-\phi_i}{2N}P(N,\phi_c;\phi_i)
      \ .
      \label{eq: one-point probability}
   \end{split}
\end{align}
Here, the e-folding number $N$ is related to the wavenumber $k$.
At the horizon crossing of perturbations, the wavenumber $[\simeq (\mathrm {bubble~size})^{-1}]$ is given by $k=a H_{\rm inf}$.
In our definition of $N$, we set $k_0=a_iH_\mathrm{inf}$ and then,
\begin{equation}
    N = \ln \frac{k}{k_0}
    \ ,
    \label{eq: e-fold def}
\end{equation}
where we take $k_0=2.24\times 10^{-4}\,\mathrm{Mpc}^{-1}$. 
The resultant volume fractions of bubbles for $\phi_i = \phi_c -4.5H_{\rm{inf}}$, $\phi_c -5H_{\rm{inf}}$, and $\phi_c -5.5H_{\rm{inf}}$ are shown in Fig.~\ref{fig: volume fraction}.

\begin{figure}[t]
    \centering
    \includegraphics[width=.75\textwidth ]{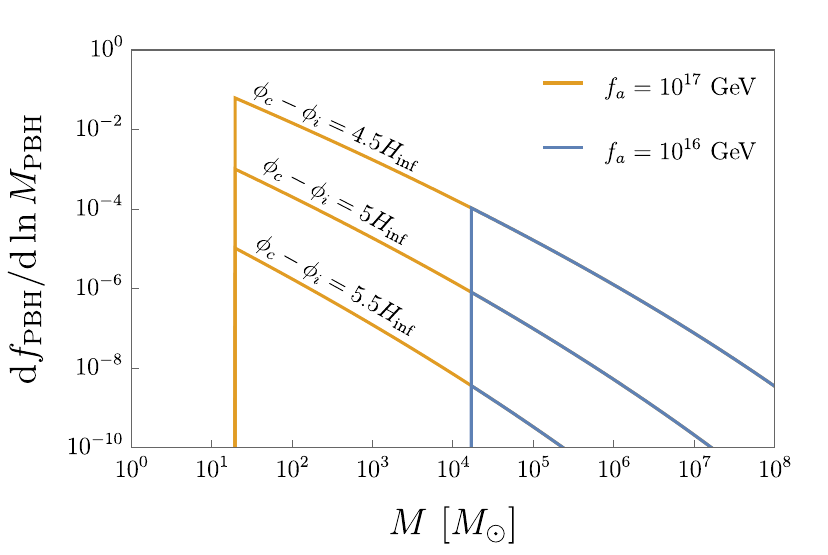}
    \caption{%
        Mass spectrum of the PBH fraction.
        While the initial condition controls the overall PBH abundance, $f_a$ controls the mass cutoff, where the PBH mass spectrum has a peak.
    }
    \label{fig: PBH fraction}
\end{figure}

The temperature when the energy density of the background radiation $\rho_\mathrm{rad}$ becomes equal to that of axions inside bubbles, $T_\mathrm{B}$, is calculated as
\begin{align}
\begin{split}
    T_\mathrm{B}
    &=
    \frac{4}{3} 
    m_{a0}
    \left.\frac{n_a}{s}\right|_\mathrm{bubble}
   \\ &=
   3.04~\mathrm{MeV}
   \left( \frac{g_{\rm osc}}{60} \right)^{-0.418}
   \left( \frac{f_a}{10^{16}~\mathrm{GeV}} \right)^{1.16}
   \\
    &\quad\times
    \left[
        1
        +0.0842\ln\left( 
            \frac{f_a}{10^{16}~\mathrm{GeV}} 
        \right)
        -0.0302\ln\left( 
            \frac{g_\mathrm{osc}}{60} 
        \right)
   \right]^{1.16}
   \ ,
\end{split}
\end{align}
where we assumed $T_\mathrm{B} <0.15\,\mathrm{GeV}$, which holds for the parameters of our interest.
At the cosmic temperature $T < T_\mathrm{B}$, axion bubbles collapse into PBHs when they reenter the horizon, which sets the lower bound on the PBH mass $M_\mathrm{PBH}$ as
\begin{align}
\begin{split}
    M_c
    &=
    1.68 \times 10^4M_{\odot}
    \left(\frac{g_f}{10}\right)^{-\frac{1}{2}}
    \left( \frac{g_{\rm osc}}{60} \right)^{0.836}
    \left( \frac{f_a}{10^{16}~\mathrm{GeV}} \right)^{-2.33}
    \\
    &\quad\times
    \left[
        1
        +0.0842\ln\left( 
            \frac{f_a}{10^{16}~\mathrm{GeV}} 
        \right)
        -0.0302\ln\left( 
            \frac{g_\mathrm{osc}}{60} 
        \right)
   \right]^{-2.33}
   \ ,
\end{split}
\end{align}
where $g_f$ is the effective relativistic degrees of freedom at the PBH formation.
Here, we use the fact that the PBH mass is equal to the background horizon mass when the axion bubbles enter the horizon~\cite{Kopp:2010sh, Carr:2014pga}. 
Note that this means that the mass of a PBH is not given by the energy density integrated over the region that collapses into the PBH.
Rather, it is given by a mass contained in the Hubble radius determined by the background energy density.
Then, we obtain the relation between $M_{\rm PBH}$ and $k$ as
\begin{align}
    k =4.55\times 10^6~\mathrm{Mpc^{-1}}
    \left(\frac{g_f}{10}\right)^{-\frac{1}{12}}
    \left(\frac{M_\odot}{M_{\mathrm{PBH}}}\right)^\frac{1}{2}
    \ .
    \label{eq: relation k and M}
\end{align}
The critical scale $k_c$ corresponding to $M_c$ is given by
\begin{align}
\begin{split}
    k_c
    &=
    3.51 \times 10^4\,\mathrm{Mpc}^{-1}
    \left(\frac{g_f}{10}\right)^{\frac{1}{6}}
    \left( \frac{g_{\rm osc}}{60} \right)^{-0.418}
   \left( \frac{f_a}{10^{16}~\mathrm{GeV}} \right)^{1.16}
   \\
    &\quad\times
    \left[
        1
        +0.0842\ln\left( 
            \frac{f_a}{10^{16}~\mathrm{GeV}} 
        \right)
        -0.0302\ln\left( 
            \frac{g_\mathrm{osc}}{60} 
        \right)
   \right]^{1.16}.
\end{split}
\end{align}
Now we have relations among $k$, $N$, and $M_\mathrm{PBH}$.
Thus, we can define one quantity as a function of another quantity.
For example, we will use $N(M_\mathrm{PBH})$ and $N(k)$ in the following.
The PBH fraction to dark matter $f_\mathrm{PBH}$ is obtained as
\begin{align}
   \frac{{\rm d}f_{\rm{PBH}}}{{\rm d}\ln{M_\mathrm{PBH}}} 
   &\equiv
       \frac{1}{\rho_{\rm{DM}}}
       \frac{{\rm d}\rho_{\rm{PBH}}}{{\rm d}\ln{M_\mathrm{PBH}}}
   \nonumber\\
   &=\frac{\rho_{\rm{rad}}(T_f) }{2\rho_{a,\mathrm{DM}}(T_f)} 
       \beta_{N,1} (N(M_\mathrm{PBH}); \phi_i)
       \,\Theta(M_\mathrm{PBH}-M_c)
    \nonumber \\
   &=
   \frac{3T_f}{8m_{a0}} 
   \left(\frac{n_{a, \mathrm{DM}}} {s}\right)^{-1}
       \beta_{N,1} (N(M_\mathrm{PBH}); \phi_i)
       \,\Theta(M_\mathrm{PBH}-M_c)
       \ ,
   \label{eq: PBH fraction}
\end{align}
where $\rho_{a, \mathrm{DM}}$ and $n_{a, \mathrm{DM}}$ are the energy density and number density of the background axion responsible for dark matter, respectively, $T_f$ is the temperature at the PBH formation, $\Theta(x)$ represents a step function, and we used $|{\rm d} \ln M_\mathrm{PBH}| = 2|\mathrm{d} \ln{k}| = 2 |\mathrm{d}N|$. 
We show the mass spectra of PBHs for $f_a = 10^{16}$\,GeV and $10^{17}$\,GeV in Fig.~\ref{fig: PBH fraction}, where each spectrum has a peak at the cutoff mass $M_c$. 
We approximate the spectrum of PBHs as monochromatic ones, i.e. $\mathrm{d} f_\mathrm{PBH}/\mathrm{d} \ln M \propto \delta(M-M_c)$ in the following. Then we evaluate $f_\mathrm{PBH}$ by ${\rm{d}}f_\mathrm{PBH}/{\rm{d}}\ln{M_\mathrm{PBH}}$ in Eq.~\eqref{eq: PBH fraction} at the peak mass.
Here and hereafter, we use $g_{\rm osc} = 60$ and $10$ for $f_a = 10^{16}~\mathrm{GeV}$ and $10^{17}~\mathrm{GeV}$, respectively, and $g_f=10$ independent of the decay constant.

On the other hand, axion bubbles  smaller than $k_c^{-1}$ do not form PBHs but axion miniclusters.
The mass of axion miniclusters is determined by the axion density in the bubbles, and its formula depends on whether the horizon entry takes place before the onset of axion oscillations or not. 
Here, we introduce the scale $k_{\rm osc}$ that reenters the horizon when the axion starts oscillation ($T \simeq T_{\rm osc}$). 
The miniclusters with scale $k < k_{\rm osc}$ are formed after the axion oscillation, and their fraction to dark matter $f_\mathrm{AMC}$ is estimated as 
\begin{equation}
   \frac{{\rm d}f_{\rm{AMC}}}{{\rm d}\ln k} 
   \equiv
   \frac{1}{\rho_{\mathrm{DM}}}\frac{{\rm d}\rho_{\mathrm{AMC}}(k)}{{\rm d}\ln{k}}
   =\frac{\rho_{\mathrm{in}}(k)}{\rho_{a,\mathrm{DM}}}
   \beta_{N1} (N(k);\phi_i)
   =
   \frac{\tilde{n}_a/s|_{\phi_\mathrm{osc}=\pi f_a-\phi^{(0)}_{\mathrm{min}}}}{n_a/s|_{\phi_\mathrm{osc}=\phi^{(0)}_{\mathrm{min}}}} 
   \beta_{N1} (N(k);\phi_i)
   \ ,
   \label{eq: bubble fraction}
\end{equation}
where $\rho_{\rm AMC}$ is the energy density of the axion miniclusters averaged over the whole universe, $\rho_{\rm in}$ is that inside the axion miniclusters, and $\tilde{n}_a$ is defined by changing $m_a(T_\mathrm{osc})\rightarrow\sqrt{m_a^2 (T_{\mathrm{osc}}) + k^2/a^2(T_{\rm osc})}$ in the definition of $n_a$, Eq.~(\ref{eq: axion number density}).
We assume that the mass of axion miniclusters is equal to the total axion mass inside axion bubbles when they enter the horizon.
Then, we obtain the relation between $M_\mathrm{AMC}$ and $k$ as
\begin{align}
\begin{split}
    M_\mathrm{AMC}(k) 
    &=2.04\times 10^{-2}~M_{\odot}
    \left(\frac{g_\mathrm{osc}}{60}\right)^{-0.418}
    \left(\frac{f_a}{10^{16}~\mathrm{GeV}}\right)^{1.16}
    \left(\frac{k_\mathrm{osc}}{k}\right)^3
    \\
    &\quad\times
    \left[
        1
        +0.0842\ln\left( 
            \frac{f_a}{10^{16}~\mathrm{GeV}} 
        \right)
        -0.0302\ln\left( 
            \frac{g_\mathrm{osc}}{60} 
        \right)
   \right]^{1.16}
    \ .
    \label{eq: relation k and AMC mass}
\end{split}
\end{align}
For miniclusters with scale $k> k_{\rm osc}$, $f_\mathrm{AMC}$ is given by
\begin{align}
    \frac{{\rm d}f_{\mathrm{AMC}}}{{\rm d}\ln k} 
    &=
    \frac{\tilde{n}_a/s|_{\phi_\mathrm{osc} = \phi_{\mathrm{dec}}}}{n_a/s|_{\phi_\mathrm{osc} = \phi_{\mathrm{DM}}}} 
    \beta_{N,1} (N(k);\phi_i)
    \ ,
    \\
    \phi_{\rm dec}
    &=(\pi f_a-\phi_{\mathrm{DM}})
        \left(\frac{m_a (T_{\mathrm{osc}}) }{3H_k}\right)^{\frac{1}{2}},
   \label{eq: bubble fraction2}
\end{align}
where $H_k$ is the Hubble parameter when the axion bubbles with the scale $k$ reenter the horizon.
Here, we take into account the decrease of $\phi$ from the horizon crossing to the onset of axion oscillations~\cite{Kitajima:2020kig}.

\begin{figure}[t]
    \centering
    \subfigure[$f_a=10^{16}~\rm{GeV}$]{%
       \includegraphics[width=.5\textwidth ]{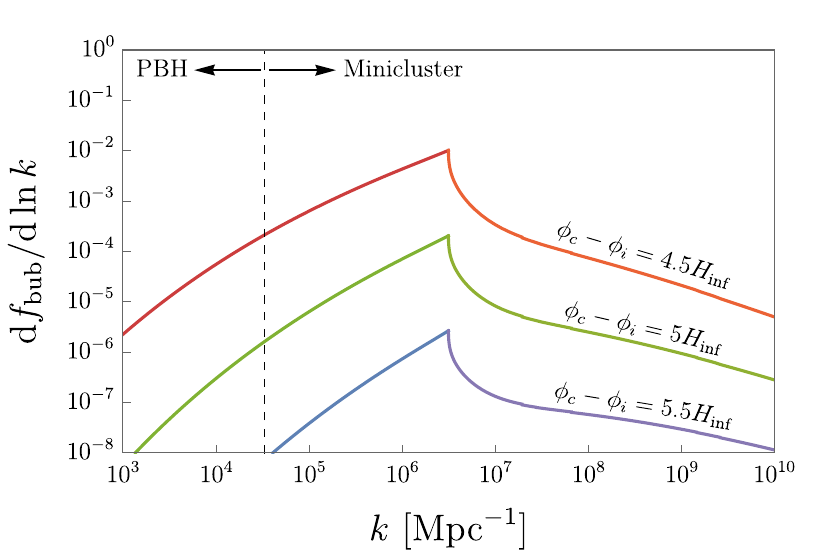}}%
    \subfigure[$f_a=10^{17}~\rm{GeV}$]{%
       \includegraphics[width=.5\textwidth ]{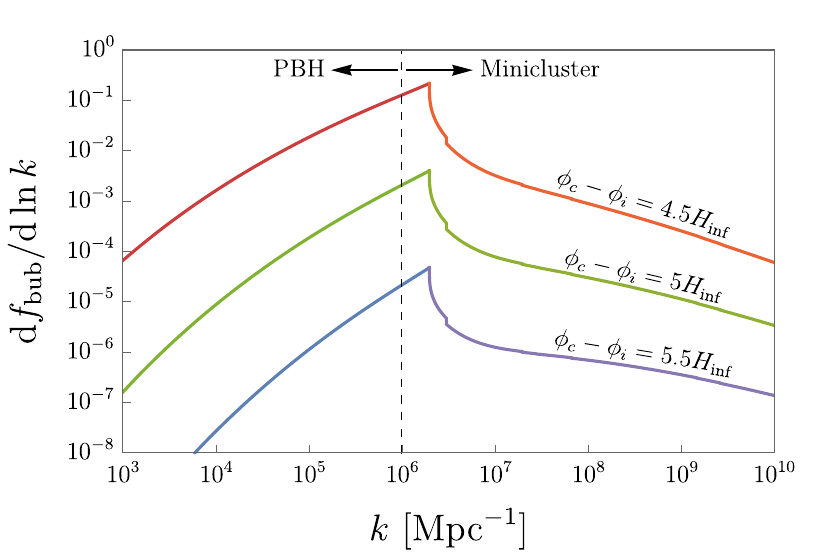}}%
    \caption{%
        Energy density fraction of axion bubbles for $f_a = 10^{16}$\,GeV (left panel) and $f_a = 10^{17}$\,GeV (right panel).
    }
    \label{fig: bubble fraction}
\end{figure}

We show the energy spectrum of axion bubbles in Fig.~\ref{fig: bubble fraction}, where $f_\mathrm{bub}$ is the global energy fraction of axion bubbles. 
In this figure, we take into account the dependence of the relativistic degrees of freedom on the temperature at the horizon reentry of mode $k$.
The energy density of axion miniclusters has a peak at $k=k_{\rm osc}$. 
Due to the Eq.~(\ref{eq: relation k and AMC mass}), we approximate the mass spectrum of axion miniclusters as monochromatic ones, i.e. $\mathrm{d} f_\mathrm{AMC}/\mathrm{d} \ln M \propto \delta(M-M_\mathrm{AMC}(k_\mathrm{osc}))$ in the following. Then we evaluate $f_\mathrm{AMC}$ by ${\rm{d}}f_\mathrm{AMC}/{\rm{d}}\ln{M_\mathrm{AMC}}$ at the peak mass using Eqs.~\eqref{eq: bubble fraction} and \eqref{eq: relation k and AMC mass}.

\section{Reduced PBH correlation function}
\label{sec: correlation_func}
The PBH clustering is characterized by the correlation function of the PBH density, $\langle \delta_{\rm PBH}(\bm{x})\delta_{\rm PBH}(\bm{y})\rangle$, with $\delta_{\rm PBH}(\bm{x}) = (\rho_{\rm PBH}(\bm{x})-\bar\rho_{\rm PBH})/\bar\rho_{\rm PBH}$, $\bar\rho_{\rm PBH}$ being the spatial average of the PBH density, $\rho_\mathrm{PBH}(\bm{x})$, over the whole universe.
While the two-point correlation function with $\bm{x} = \bm{y}$ corresponds to the Poisson fluctuations, that with $\bm{x} \neq \bm{y}$, $\xi(|\bm{x}-\bm{y}|)$, reflects the effect of the clustering.
Here, the angular bracket denotes the ensemble average.
$\xi(L)$ is called the reduced PBH correlation function and obtained as~\cite{Kawasaki:2021zir}
\begin{align}
   \xi(L)
   &\equiv
   \frac{
        \int\int {\rm{d}}\ln{M_i}{\rm{d}}\ln{M_j}
            \rho_{\rm{rad}}(M_i)
            \rho_{\rm{rad}}(M_j)
            \beta_{N,2}(N(M_i),N(M_j),N(2\pi/L);\phi_i)/4}
    {\left(\int {\rm{d}}\ln{M}
        \rho_{\rm{rad}}(M)
        \beta_{N,1}(N(M);\phi_i)/2\right)^2}-1
    \ ,
\end{align}
where $\rho_\mathrm{rad}(M)$ is the radiation energy density when the scale corresponding to PBHs with a mass of $M$ enters the horizon, and $\beta_{N,2}$ is the formation probability of axion bubbles located at two points, $x$ and $y$, whose comoving distance is $L$.
Note that $|{\rm{d}}\ln{M}_\mathrm{PBH}|=2|{\rm{d}}\ln{k}|$. 
Suppose that the sizes of the two bubbles located at $x$ and $y$ are $2\pi k_x^{-1}$ and $2\pi k_y^{-1}$, respectively.
Following the discussion in Ref.~\cite{Kawasaki:2021zir}, $\beta_{N,2}$ is evaluated as
\begin{equation}
   \begin{split}
       \beta_{N,2}(N_x,N_y,N_L;\phi_i)
       &\equiv
       \frac{\partial^2}{\partial N_x \partial N_y}
       \int {\rm{d}}\phi_L \, P(N_L,\phi_L;\phi_i)
       \\
       &\quad
       \times \int {\rm{d}}\phi_x \, P(N_x-N_L,\phi_x;\phi_L)
       \Theta(\phi_x-\phi_c)
       \\
       &\quad
       \times \int {\rm{d}}\phi_y \, 
       P(N_y-N_L,\phi_y;\phi_L) \Theta(\phi_y-\phi_c)\ ,
    \end{split}
   \label{eq: two-point probability origin}
\end{equation}
where $N_L$, $N_x$, and $N_y$ ($\phi_L$, $\phi_x$, and $\phi_y$) are e-folding numbers (axion field values) when the scales $L$, $2\pi k_x^{-1}$, and $2\pi k_y^{-1}$ exit the horizon during inflation, respectively.
The axion fields at the two points, $x$ and $y$, evolve in the same way as long as they are inside the same Hubble horizon.
When their separation becomes the horizon size, which corresponds to $N=N_L$, the field values at $x$ and $y$ start to evolve independently from $\phi_L$. 
Note that $\phi_L$ can be regarded as a background value over the scale $L$ (before the appearance of the axion potential) but it spatially fluctuates beyond this scale.
$N_x$ and $N_y$ can be translated into the PBH mass using Eq.~\eqref{eq: relation k and M} or the mass of axion miniclusters using Eq.~(\ref{eq: relation k and AMC mass}).

\begin{figure}[t]
    \centering
    \includegraphics[width=.625\textwidth]{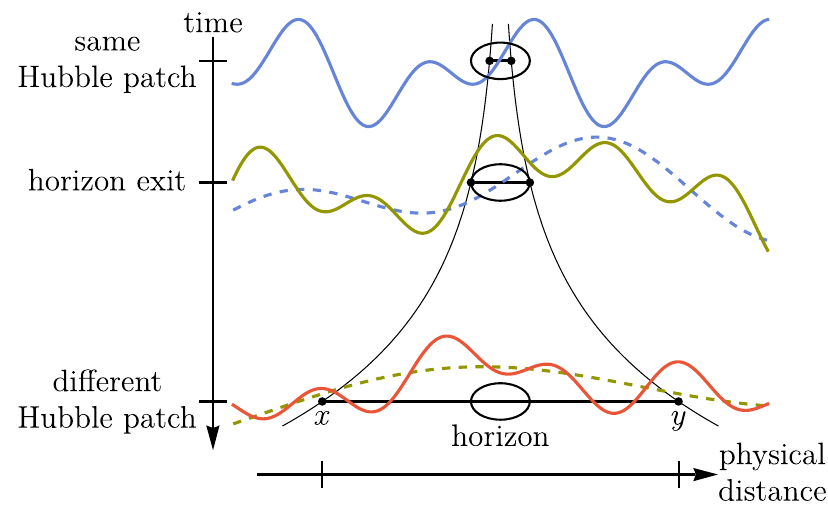}
    \caption{This shows a schematic image of the calculation of $\beta_{N,2}$ during the inflation.
    In the early stage of inflation, two spatial points, $x$ and $y$, reside in the same Hubble patch (circle) and share the superhorizon fluctuations in $\phi$ (blue line). Fluctuations larger than our interesting scale (green line) are continuously added to the superhorizon fluctuations (dashed blue line) until the horizon exit of the two points.
    After then, the physical distance between the two points exceeds the horizon scale, and they are in different Hubble patches.
    The coarse-grained fields around $x$ and $y$ include the common contribution (dashed green line) and independent contributions (red line).
    As a result, the field values at two separated points are correlated, which leads to the bubble clustering
    (see also Fig.~\ref{fig: what causes clusterings}).}
    \label{fig: schematic view} 
\end{figure}

We show the schematic image of our calculation in Fig.~\ref{fig: schematic view}.
The solid line shows spatial fluctuations of the field value smoothed over the horizon scale at each time, and the dashed line shows the fluctuations which exited the horizon at the former time.
The larger $\phi_L$ is, the more bubbles are formed in the region with comoving size $L$.
Since $\phi_L$ spatially fluctuates beyond the scale $L$, the number of bubbles fluctuates accordingly.
As a result, the bubbles form clusters.

\begin{figure}[t]
    \centering
    \subfigure[$f_a=10^{16}~\rm{GeV}$]{%
       \includegraphics[width=.5\textwidth ]{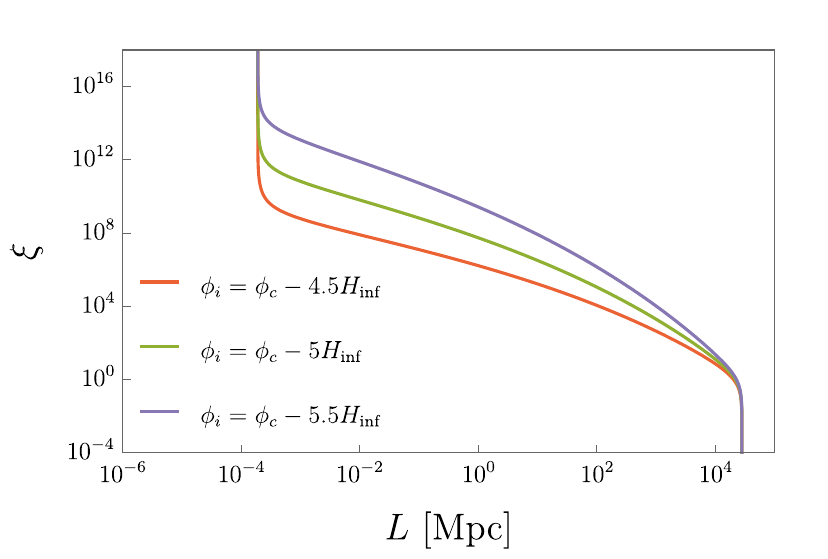}}%
    \subfigure[$f_a=10^{17}~\rm{GeV}$]{%
       \includegraphics[width=.5\textwidth ]{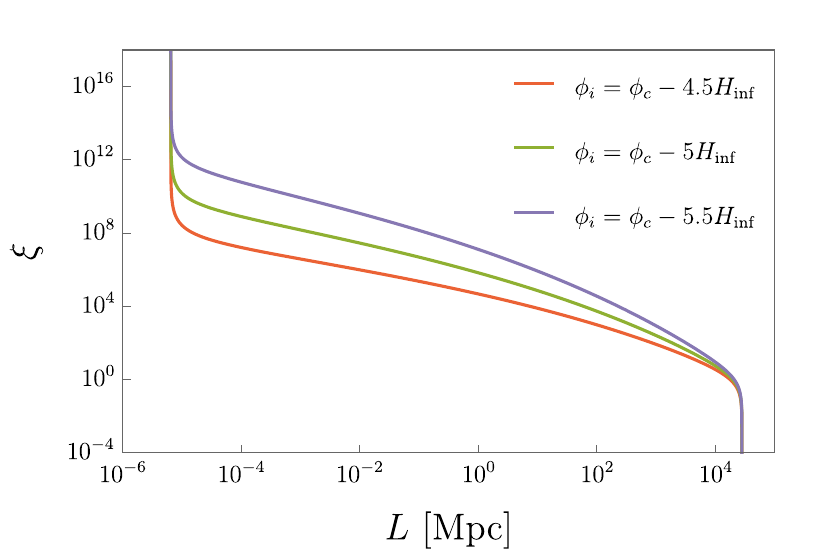}}%
    \caption{%
        Reduced PBH correlation functions for $f_a = 10^{16}$\,GeV (left panel) and $f_a = 10^{17}$\,GeV (right panel).
        They diverge at the bubble scale and vanish at the scale of the observable universe.
    }
    \label{fig: correlation}
\end{figure}

As mentioned above, we consider a monochromatic mass in the following.
Thus, we focus on $N_x = N_y =N$.
By substituting Eq.~\eqref{eq: probability density} into Eq.~\eqref{eq: two-point probability origin}, we obtain $\beta_{N,2}$ analytically as
\begin{align}
   &\beta_{N,2}(N, N, N_L;\phi_i)
    =\frac{H_\mathrm{inf}^4}{128\pi^5\sigma^5(N-N_L)}\exp{\left[
               -\frac{(\phi_i-\phi_c)^2}
                     {\sigma^2(N-N_L)+2\sigma^2(N_L)}
                \right]}
    \nonumber \\
    &\quad \quad\times \Bigg[\sigma^2(N_L)\sigma^2(N-N_L)
               \left(
                   \sigma^2(N-N_L)+2\sigma^2(N_L)
               \right)^{-\frac{3}{2}}
    \nonumber \\
    &\quad\quad+\left(
           1-\frac{2\sigma^2(N_L)}{\sigma^2(N-N_L)+2\sigma^2(N_L)}
          \right)^2
          (\phi_i-\phi_c)^2
          \left(
              \sigma^2(N-N_L)+2\sigma^2(N_L)
          \right)^{-\frac{1}{2}}
        \Bigg]
        \ .
   \label{eq: two-point probability}
\end{align}
Here $N_L$ can be translated into $L$ using Eq.~(\ref{eq: e-fold def}) with $k_L=2\pi/L$.
Then, the reduced PBH correlation function is obtained as
\begin{align}
   \xi(L)
   =
    \frac{\beta_{N,2}(N(M),N(M),N(2\pi/L);\phi_i)}{\beta_{N,1}^2(N(M);\phi_i)}-1
    \ .
   \label{eq: correlation}
\end{align}
We show the correlation function $\xi(L)$ for $f_a = 10^{16}$\,GeV and $10^{17}$\,GeV in Fig.~\ref{fig: correlation}.
For smaller $L$, the correlation function increases because the two points reside in the same Hubble patch for a longer time during inflation.
The correlation function diverges when the separation of the two points becomes equal to the bubble scale, $L = 2 \pi k_\mathrm{PBH}^{-1}$. 
Since this bubble scale is negligible compared to the CMB scale, which is of our interest, we shift the divergent point to the origin in the following. 
In other words, we redefine $\xi(L-2\pi k_{\rm{PBH}}^{-1})$ as a new correlation function $\xi(L)$.
The correlation function vanishes at the size of the observable universe.
This is because the probability of PBH formation at two points $\beta_{N,2}$ is equal to a product of two independent probabilities of PBH formation $\beta_{N,1}^2$ at the edge of the observable universe.

\section{Isocurvature perturbation}
\label{sec: isocurvature}

In this section, we evaluate the isocurvature perturbations of the PBHs and axion miniclusters and derive the constraints on the scenario. 
Here, we mainly explain the evaluation for PBHs because the procedure is the same for the axion miniclusters.

First, we consider the isocurvature perturbations from PBHs.
The power spectrum of PBH density fluctuations can be written as
\begin{equation}
   P_{\rm{PBH}}(k)
   =
   \int {\rm d}^3x~e^{-i \bm{k}\cdot \bm{x}}
   \left< \delta_{\rm{PBH}}(0) \delta_{\rm{PBH}}(\bm{x}) \right>
   =
   P_{\rm{Poisson}}(k) + P_{\xi}(k)
   \ ,
\end{equation}
where the contributions from the Poisson fluctuation and the PBH clustering are given by  
\begin{align}
    P_{\rm{Poisson}}(k)
    &\equiv
    \frac{1}{\bar{n}_{\mathrm{PBH}}}
    \frac{\overline{M_\mathrm{PBH}^2}}{(\overline{M_\mathrm{PBH}})^2}
    \ ,
    \\
    P_{\xi}(k)
    &\equiv
    \int {\rm d}^3x~\xi(x)e^{-i
    \bm{k}\cdot \bm{x}}
    \label{eq: P_xi integration}
    =
    4\pi \int {\rm d}r~r^2\xi(r)\frac{\sin{kr}}{kr}
    \ ,
\end{align}
respectively.
Here,
$\bar{n}_\mathrm{PBH}$ is the spatially averaged PBH number density, and the bars on $M_{\rm PBH}$ and $M_{\rm PBH}^2$ mean the average over the PBH mass distribution. 

\begin{figure}[t]
    \centering
    \subfigure[$f_a=10^{16}~\mathrm{GeV}~(\phi_c-\phi_i = 5.5 H_{\rm{inf}})$]{%
       \includegraphics[width=.48\textwidth ]{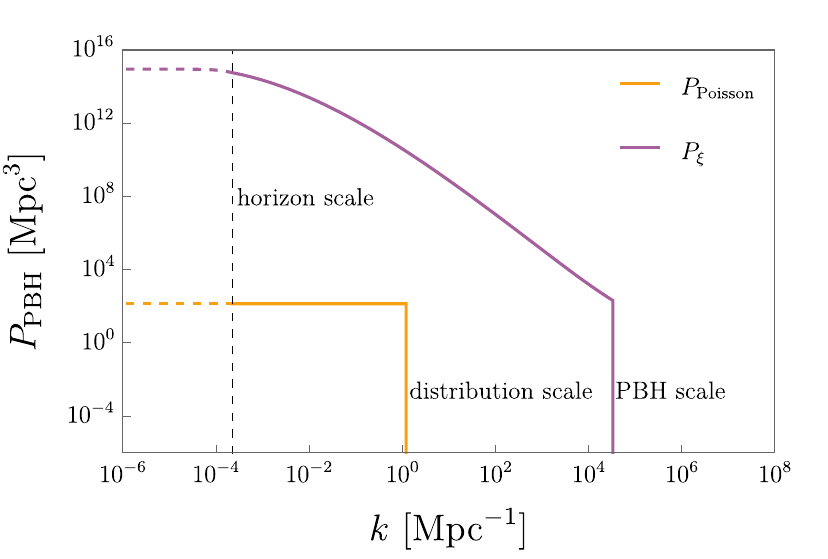}}%
    \subfigure[$f_a=10^{17}~\mathrm{GeV}~(\phi_c-\phi_i = 6 H_{\rm{inf}})$]{%
       \includegraphics[width=.48\textwidth ]{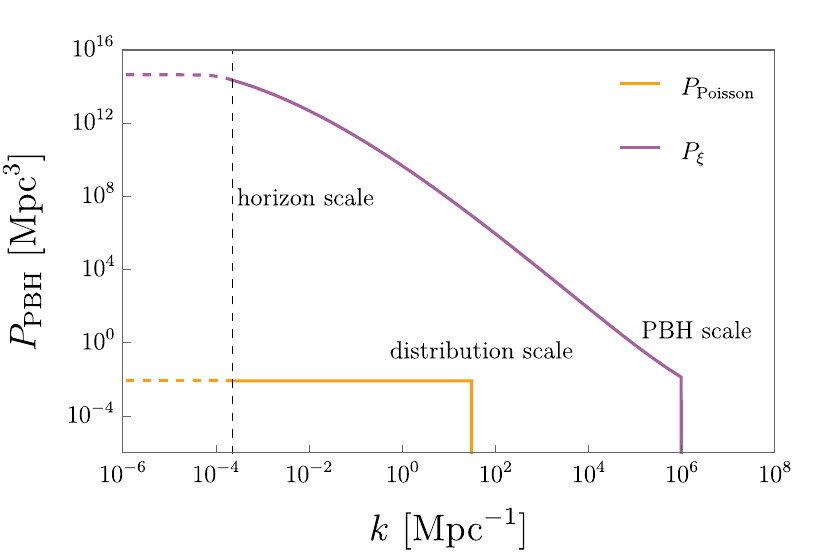}}%
    \caption{%
        Power spectrum of density fluctuations of PBHs for $f_a = 10^{16}$\,GeV and $\phi_c - \phi_i = 5.5 H_\mathrm{inf}$ (left panel) and $f_a = 10^{17}$\,GeV and $\phi_c - \phi_i = 6 H_\mathrm{inf}$ (right panel). 
    }
    \label{fig: power spectrum for k}
\end{figure}

Now, we approximate the PBH mass distribution by a monochromatic one and then obtain $P_{\rm{Poisson}}(k)=\bar{n}_{\rm{PBH}}^{-1}$. 
Because the Poisson fluctuations are caused by the randomness of the spatial PBH distribution, we set a cutoff of the Poisson fluctuations at the scale where the expectation value of the PBH number in the corresponding volume becomes unity.
This corresponds to the scale at which the fluctuation becomes larger than the expectation value and thus the PBH density fluctuation becomes nonlinear~\cite{Gong:2018sos}.
Moreover, we evaluate $P_\xi$ from Eq.~\eqref{eq: correlation}.
Because PBHs are formed from axion bubbles in this scenario, the fluctuations from PBH clustering only make sense on scales larger than the bubble scales.
Thus, we show the PBH power spectrum between the current horizon scale and the bubble scale in Fig.~\ref{fig: power spectrum for k}.
It is seen that the density perturbations caused by the clustering of PBHs are much larger than those expected from the Poisson statistics.
Since the energy density of axion miniclusters also has a peak as shown in Fig.~\ref{fig: bubble fraction}, we obtain $P_\xi$ for axion miniclusters, $P_\mathrm{AMC}$, by the same procedure.

\begin{figure}[t]
    \centering
    \subfigure[$f_a=10^{16}~\rm{GeV}$]{%
       \includegraphics[width=.48\textwidth ]{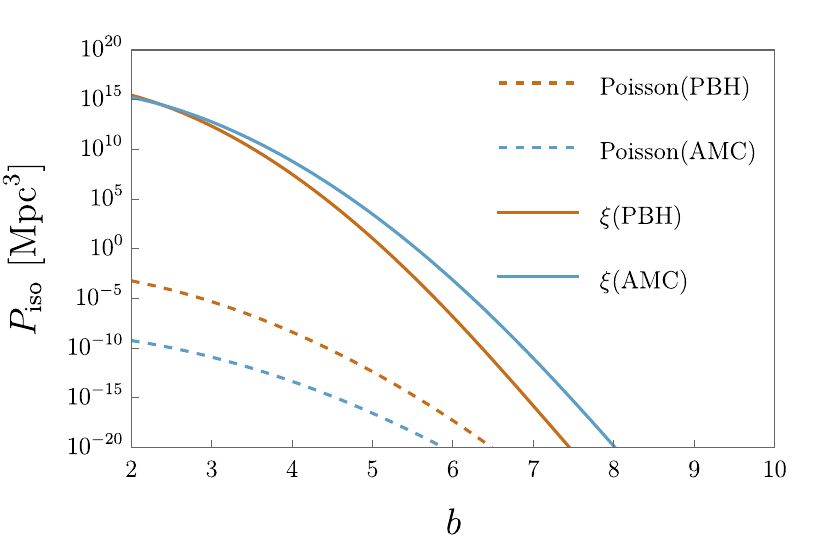}}%
    \subfigure[$f_a=10^{17}~\rm{GeV}$]{%
       \includegraphics[width=.48\textwidth ]{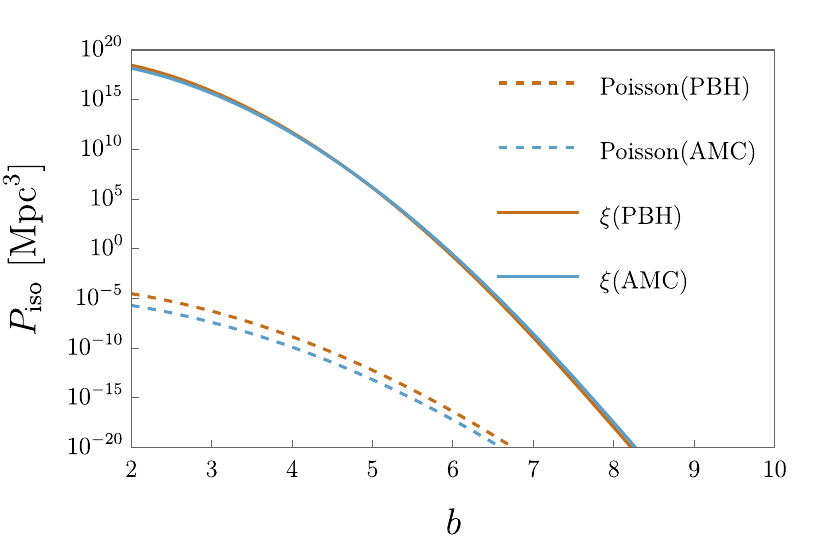}}%
    \caption{%
        Amplitudes of the power spectrum of the isocurvature perturbations from the PBHs and axion miniclusters on the CMB scale for $f_a = 10^{16}$\,GeV (left panel) and $f_a = 10^{17}$\,GeV (right panel). 
        We parameterize the initial condition with $b$ defined by $\phi_c-\phi_i \equiv b H_{\rm{inf}}$.
    }
    \label{fig: power spectrum}
\end{figure}

The perturbations of the PBHs contribute to the CDM isocurvature perturbations.
Th e power spectrum of the isocurvature perturbations is calculated as
\begin{equation}
   P_{\rm{iso}}(k)=
   \left(\frac{\rho_{\rm{PBH}}}{\rho_{\rm{DM}}}\right)^2P_{\rm{PBH}}(k)
   \ .
\end{equation}
The amplitude of the isocurvature perturbation on the CMB scale is strongly constrained by the Planck results as~\cite{Planck:2018jri}
\begin{equation}
   \beta_{\rm{iso}}(k_\mathrm{CMB})
   \equiv
   \frac{ P_{\rm{iso}}(k_\mathrm{CMB}) }{ P_{\rm{iso}}(k_\mathrm{CMB})+P_{\mathcal{R}}(k_\mathrm{CMB}) }
   <
   0.036 \ ,
   \label{eq: obs const}
\end{equation}
where $\mathcal{P}_{\mathcal{R}}(k_\mathrm{CMB})= (k_\mathrm{CMB}^3/2\pi^2)P_\mathcal{R} \simeq 2\times 10^{-9}$ is the power spectrum of the curvature perturbations at $k_{\rm CMB} = 0.002$~Mpc$^{-1}$~\cite{Planck:2018vyg}.
Here, this observational constraint (\ref{eq: obs const}) is applicable not only to PBHs but also to axion miniclusters.
Although we follow the discussion in Ref.~\cite{Kawasaki:2021zir} so far, the isocurvature perturbation from axion miniclusters is unique to the axion bubble scenario.
The amplitude of the power spectrum of the isocurvature perturbations from the PBHs and axion miniclusters on the CMB scale is shown in Fig.~\ref{fig: power spectrum}, where we parameterize the initial condition with $b$ defined by $\phi_c-\phi_i \equiv b H_{\rm{inf}}$ instead of $\phi_i$. 

\begin{figure}[t]
    \centering
    \includegraphics[width=.75\textwidth ]{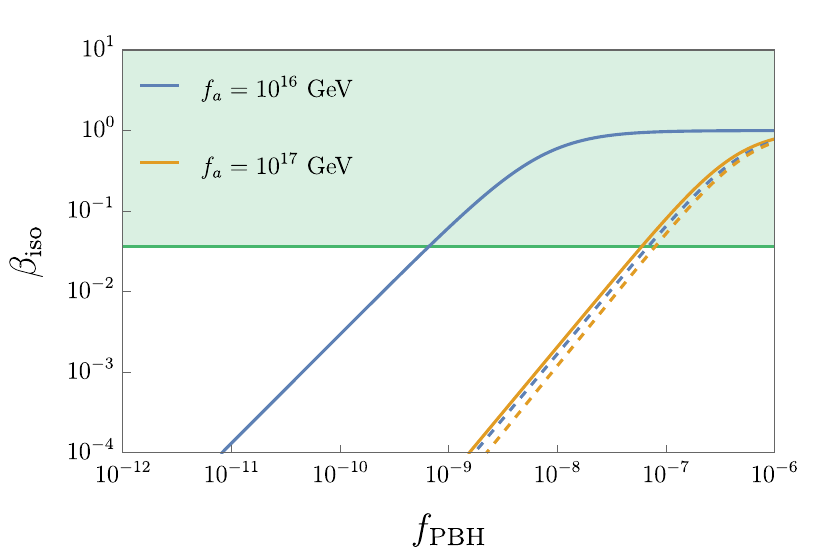}
    \caption{%
    Isocurvature perturbations from PBHs (dashed) and axion miniclusters (solid) for $f_a = 10^{16}$\,GeV and $10^{17}$\,GeV.
    The green-shaded region is constrained by the CMB observations, which set the upper bounds on $f_\mathrm{PBH}$.
    }
    \label{fig: constraints}
\end{figure}

Since $f_\mathrm{PBH}$ is determined by $b$ for a given $f_a$, we can predict $\beta_\mathrm{iso}$ as a function of $f_\mathrm{PBH}$, which is shown in Fig.~\ref{fig: constraints}.
From the figure, we obtain the constraints on the PBH fraction $f_{\rm{PBH}}$ as
\begin{equation}
    f_\mathrm{PBH} ~\lesssim ~ 
        \begin{cases}
          7\times 10^{-8} ~~& \mathrm{from~PBHs} \\
          7 \times 10^{-10}& \mathrm{from~axion~miniclusters} 
        \end{cases}
        ~~\mathrm{for}~~ f_a =10^{16}~\rm{GeV},
\end{equation}
and
\begin{equation}
    f_\mathrm{PBH} ~\lesssim ~ 
        \begin{cases}
          8\times 10^{-8} ~~& \mathrm{from~PBHs} \\
          6 \times 10^{-8}& \mathrm{from~axion~miniclusters} 
        \end{cases}
        ~~\mathrm{for}~~ f_a =10^{17}~\rm{GeV}.
\end{equation}
We find that the constraints from axion miniclusters are more stringent than those from PBHs. 

\section{Conclusion and discussion}
\label{sec: summary}

In this paper, we have studied the clustering of PBHs and axion miniclusters produced in the axion bubble model~\cite{Kitajima:2020kig}.
To investigate the clustering, we evaluated the two-point correlation function of PBHs and miniclusters and the isocurvature perturbations, from which we derived stringent constraints on the model parameters and the PBH abundance.

For the axion decay constant $f_a = 10^{17}~\mathrm{GeV}$, which corresponds to the PBH mass $\mathcal{O}(10)M_\odot$ motivated by the LIGO/Virgo events, the PBH fraction in dark matter is constrained as $f_{\rm{PBH}}\lesssim 6\times 10^{-8}$.
Thus, the PBH abundance should be much smaller than the typically required abundance for the LIGO/Virgo events, $f_\mathrm{PBH} \simeq 10^{-3}\,$--$\,10^{-2}$~\cite{Sasaki:2016jop}.
However, this required abundance is evaluated neglecting the clustering effects on the PBH mergers.
In fact, the evaluation of the merger rate is significantly affected by the strong clustering as shown in Ref.~\cite{Kawasaki:2021zir} although precise estimation is still difficult because of the three-body problem.
Thus, we need to develop a method to treat PBH mergers in multi-body systems to derive a definite conclusion.

On the other hand, in the case of $f_a = 10^{16}~\mathrm{GeV}$, the axion bubble model can explain the seeds of SMBHs.
Our constraint $f_{\rm{PBH}}\lesssim 7\times10^{-10}$ for $f_a = 10^{16}~\mathrm{GeV}$ is more severe than the CMB constraint for $M_\mathrm{PBH} \leq 3 \times 10^4 M_\odot$ from PBH accretion~\cite{Serpico:2020ehh}.
The mass fraction of SMBHs inferred from the observation~\cite{Willott:2010yu} is $f_{\rm{SMBH}}\simeq 3\times 10^{-9}$ for $M_\mathrm{SMBH} > 10^6 M_\odot$.
In our model PBHs with mass $\sim 10^4 M_\odot$ grow to SMBHs by accretion and merger.
Thus, if the mass of PBHs increases by more than a factor of $\mathcal{O}(10)$ due to accretion, our constraint is consistent with the observation.
Thus, the axion bubble scenario can explain the seeds of the SMBHs consistently with the isocurvature constraint.

The constraints we obtained come from the large-scale isocurvature perturbations.
Thus, the constraints can be relaxed if the spectrum of the isocurvature perturbations is blue-tilted.
The blue-tilted spectrum is realized by considering that the radial component of the PQ field rolls down from a large value during the early stage of inflation~\cite{Kasuya:2009up}. 
Then, quantum diffusion in the phase direction on larger scales is suppressed because the decrease of the radial component results in a contraction in the phase direction.
Consequently, we expect that the two-point correlations on larger scales and, correspondingly, the isocurvature perturbations on larger scales become much smaller.

Finally, we comment on a complementary constraint on PBHs explaining the seeds of the SMBHs.
In the type of PBH formation models we have studied, the PBH clustering induces angular correlations~\cite{Shinohara:2021psq}. 
From a comparison of the theoretical prediction and the observational data of the angular correlation, the PBH abundance is severely constrained~\cite{Shinohara:2023wjd}. 

\begin{acknowledgments}
This work was supported by JSPS KAKENHI Grant Nos. 20H05851(M.K., N.K., and F.T.), 21K03567(M.K.), 20J20248 (K.M.), 23KJ0088 (K.M.), and 20H01894 (F.T. and N.K.), 21H01078 (N.K.), 21KK0050 (N.K.), JSPS Core-to-Core Program (grant number: JPJSCCA20200002) (F.T.), JST SPRING (grant number: JPMJSP2108) (K.K.).
K.M. was supported by the Program of Excellence in Photon Science.
K.K. was supported by the Spring GX program.
This article is based upon work from COST Action COSMIC WISPers CA21106, supported by COST (European Cooperation in Science and Technology).
\end{acknowledgments}

\bibliographystyle{JHEP}
\bibliography{Ref}

\end{document}